\documentclass[aps,twocolumn,showpacs,preprintnumbers,prl,amsmath,amssymb,amsfonts,superscriptaddress,floatfix]{revtex4-1}

\usepackage[T1]{fontenc}
\usepackage[latin1]{inputenc}
\usepackage{graphics}
\usepackage{color}

\usepackage{amssymb}
\usepackage{amsmath}
\usepackage{overpic}
\usepackage{epstopdf}
\usepackage{bm}
\usepackage{graphicx}
\usepackage{subfigure}

\newcommand{\be}{\begin{equation}}
\newcommand{\ee}{\end{equation}}
\newcommand{\bea}{\begin{eqnarray}}
\newcommand{\eea}{\end{eqnarray}}

\def\s{\sigma}

\def\br{{\bf r}}

\def\lb{\label}
\def\pref#1{(\ref{#1})}

\newcount\bozza \bozza=0
\ifnum\bozza=1
\newdimen\shift \shift=-2truecm
\def\lb#1{%
{\label{#1}\rlap{\kern\shift{$\scriptstyle#1$}}}}
\else\def\lb#1{\label{#1}} \fi

\begin{document}

\title{Broadening of the Berezinskii-Kosterlitz-Thouless transition by correlated disorder}

\author{I. Maccari}
\affiliation{ISC-CNR and Dept. of Physics, Sapienza University of Rome,
  P.le A. Moro 5, 00185, Rome, Italy}
\author{L.~Benfatto} 
\affiliation{ISC-CNR and Dept. of Physics, Sapienza University of Rome,
  P.le A. Moro 5, 00185, Rome, Italy}
\author{C.Castellani}
\affiliation{ISC-CNR and Dept. of Physics, Sapienza University of Rome,
  P.le A. Moro 5, 00185, Rome, Italy}
\date{\today}

\begin{abstract}
The Berezinskii-Kosterlitz-Thouless (BKT) transition in two-dimensional superconductors is usually expected to be protected against disorder. However, its typical signatures in real system, like e.g. the superfluid-density jump, are often 
at odd with this expectation. Here we show that the disorder-induced granularity of the superconducting state modifies the nucleation mechanism for vortex-antivortex pairs. This leads to a considerable smearing of the universal superfluid-density jump as compared to the paradigmatic clean case, in agreement with experimental observations.
\end{abstract}


\maketitle

More than 40 years after the seminal work by Berezinskii\cite{bkt} Kosterlitz and Thouless\cite{bkt1,bkt2} the BKT transition remains one of the most fascinating examples of topological phase transitions.
Its universality class describes several phenomena ranging from the quantum metal-insulator transition in one dimension to the Columb-gas screening transition in 2D, and of course the metal-to-superfluid transition in 2D\cite{review}. As such it has been investigated in neutral superfluids, as e.g. thin He films\cite{helium4} and cold-atoms systems made of bosons\cite{dalibart_nature06} or neutral fermions\cite{murthy_prl15}, but also in quasi-two-dimensional (2D) superconductors. The latter case applies not only to thin films of 
conventional \cite{fiory_prb83,lemberger_prl00,armitage_prb07,armitage_prb11,kamlapure_apl10,mondal_bkt_prl11,goldman_prl12,yazdani_prl13}   and unconventional\cite{lemberger_natphys07,lemberger_prb12,popovic_prb16} superconductors, but also to the artificially confined 2D electron gas at the interface between two insulators in artificial heterostructures\cite{bert_prb12,bid_prb16}, or in the top-most layer of ion-gated superconducting (SC) systems\cite{iwasa_science15}.

One well known difference between neutral superfluids and charged superconductors is that in the latter case the screening supercurrents must be ineffective in order to observe the BKT physics. This happens when the Pearl\cite{Pearl} length $\Lambda=2\lambda^2/d$ exceeds the system size, where $d$ is the film thickness and $\lambda$ is the magnetic penetration depth.  Such condition is usually realized when $d$ is of order of tens of nanometers. Indeed, reducing the film thickness has also the concomitant effect to enlarge $\lambda$, inversely proportional to the superfluid density, due  to the relative increase of the disorder level. This implies that the BKT transition is found in strongly disordered systems. As observed experimentally both in thin films\cite{sacepe_11,mondal_prl11,pratap_13,noat_prb13,roditchev_natphys14} and SC heterostructures\cite{biscaras_natmat13,bid_prb16,jespersen_prb16}, disorder can also induce a 
 "granular" inhomogeneous SC state, well understood theoretically\cite{trivedi_prb01,dubi_nat07,ioffe,nandini_natphys11,seibold_prl12,lemarie_prb13} as the way out of superconductivity,  which requires phase coherence, to survive in the presence of disorder-induced charge localization. 
 
 Understanding the role of the microscopic electronic disorder on the BKT transition within SC fermionic models is an incredible task\cite{dubi_nat07,meir_epl10,nandini_natphys11,mirlin_prb15}, due mainly to the small size of systems accessible numerically. Alternatively, one can address the question directly within a proper  phase-only model. A natural option is an $XY$ model with random couplings\cite{stroud_prb00,coura_prb05,meir_prl13,trivedi_prx14}, which mimics the random Josephson-like coupling between coarse-grained neighboring SC islands. As long as the random coupling are spatially uncorrelated the Harris criterium\cite{harris}  
guarantees that disorder is irrelevant, so that for example  the expected "universal" jump\cite{nelson_prl77} of the superfluid stiffness at $T_{BKT}$ is still preserved, as confirmed by numerical simulations\cite{coura_prb05,meir_prl13}. However, this finding is at odd with experiments in disordered films of conventional superconductors\cite{fiory_prb83,lemberger_prl00,armitage_prb07,armitage_prb11,kamlapure_apl10,mondal_bkt_prl11,goldman_prl12,yazdani_prl13}    where the superfluid density jump at the transition is systematically smeared out in a rapid downturn much broader than what observed in the case of superfluid helium films\cite{helium4}. The effect is even more dramatic in ultrathin films of cuprate superconductors\cite{lemberger_prb12}, where the BKT jump is completely lost by underdoping. 

In this Letter we study the effect of disorder on the BKT transition by means of Monte Carlo simulations on a disordered 2D $XY$ model. 
\be
\lb{hamrtf}
H=-\sum_{ij}J_{ij}\cos(\theta_i-\theta_j)
\ee
where $\theta_i$ is the angular variable for the 2D (planar) spins, or equivalently the SC phase in the mapping to a SC problem. The spatial arrangement of the couplings $J_{ij}$ between neighbouring sites $i,j$ is taken to mimic both uncorrelated and spatially-correlated disorder. In the latter case we use the disordered structure generated by the mean-field solution of the (quantum) $XY$ model in random transverse field (RTF), which has been recently proven model disordered superconductors with a non-trivial granular space structure \cite{ioffe,lemarie_prb13,cea_prb14}. While for uncorrelated disorder the universal superfluid-density jump is always preserved, for the RTF case the fragmentation of the SC state at strong disorder leads to a smoothening of the BKT jump, which is symmetrically smeared out with respect to the expected transition, in strong analogy with the experimental observations in thin SC films\cite{fiory_prb83,lemberger_prl00,armitage_prb07,armitage_prb11,kamlapure_apl10,mondal_bkt_prl11,goldman_prl12,yazdani_prl13}.   This result follows from an unconventional vortex-pairs nucleation in the granular SC state. 
Thus, even though the vortex unbinding remains the driving force of the transition in 2D, the identification of the BKT signatures in real, inhomogenous systems can be more subtle than usually expected. 

To identify the effects of disorder in the model \pref{hamrtf} we compute by means of Monte Carlo simulations (see \cite{suppl} for technical details) the superfluid stiffness $J_s$, which is defined for a given, let us say  $x$, direction as
\bea
\lb{js}
J_s&=&J_d-J_p,\\
\lb{jd}
J_d&=&\frac{1}{L^2}\langle \sum_i J_{i,i+x}\cos (\theta_i-\theta_{i+x})\rangle,
\\
\lb{jp}
J_p&=&\frac{1}{TL^2}\left\langle \left(\sum_{i}J_{i,i+x}\sin (\theta_i-\theta_{i+x})\right)^2\right\rangle,
\eea
where $L$ is the size of the square lattice.
Here $J_d$ denotes the diamagnetic term, which coincides with the average kinetic energy of the system, while $J_p$ is the paramagnetic term, obtained as a current-current correlation function for the paramagnetic current $I^p_{i,i+x}=J_{i,i+x}\sin (\theta_i-\theta_{i+x})$ of the model \pref{hamrtf}. Simulations are performed at $L=128$ where in the clean case ($J_{ij}=J$) a sharp jump of the superfluid stiffness is recovered\cite{suppl}. 
\begin{figure}[h!]
\includegraphics[width=8cm,clip=true]{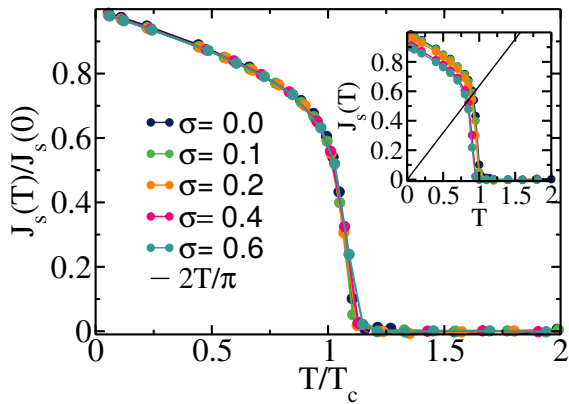}
\caption{ Superfluid stiffness as function of the temperature for a Gaussian distribution of the couplings at different values of the variance $\sigma$. In the main panel the curves have been rescaled by the value of the stiffness at $T=0$ and by $T_c$, defined as the intersection of $J_s(T)$ with the $2T/\pi$  line, shown in the inset (solid black line). }
\label{fig0}
\end{figure}

 According to the Harris criterium\cite{harris} one expects that in the presence of uncorrelated disorder the critical temperature is suppressed, but the universality of the transition and its sharpness are unaffected. Indeed, since at the BKT transition the correlation length diverges exponentially\cite{bkt2} instead than the usual power-law, the length scale set by disorder is always irrelevant for the ordering process. We checked this behavior for a paradigmatic case where the values of the local couplings $J_{ij}$ are randomly extracted from a Gaussian distribution $P_{G}(J_{ij})=\exp(-(J_{ij}-\bar J)/2\s^2)/\sqrt{2\pi\s^2}$ with increasing variance $\sigma$ and fixed average $\bar J=1$, see Fig.\ \ref{fig0}. At low temperature the primary excitations of the model \pref{hamrtf} are disordered longitudinal spin-waves, which can be well described\cite{future} by the quadratic approximation of the Hamiltonian \pref{hamrtf}, $H\approx \int d\br J(\br)(\nabla \theta(\br))^2$. By making an expansion of the local stiffness $J(\br)=\bar J+\delta J(\br)$ around its average value one can show\cite{future} that at low temperatures
 \be
 \lb{jsest}
 J_d\simeq \bar J-T/4, \quad J_p=\bar J\left[{\langle\delta J\rangle^2}/{2\bar J^2}+c(T/\bar J)^2\right]
 \ee
where $c$ is numerical constant. As a consequence, at $T=0$ disorder induces a paramagnetic suppression of the stiffness 
$J^{app}_s(T=0)\simeq \bar J\left[1-{\langle\delta J\rangle^2}/{2\bar J^2}\right]$ 
which can also be obtained\cite{stroud_prb00} by using the mapping\cite{kirkpatrick} into a random-resistor network with conductance $J_{ij}$ at each node. These results are confirmed by the simulations for Gaussian disorder shown in Fig.\ \ref{fig0}. Apart from the $T=0$ correction \pref{jsest} of the stiffness due to disorder the curves of $J_s(T)$ show a remarkable similarity, with an universal $-T/4$ depletion at low $T$ followed by a rapid downturn at the temperature $T_c$ where $J_s(T_c)=2T_c/\pi$, i.e. where the jump is predicted for the clean case at $L=\infty$\cite{nelson_prl77}. The irrelevance of disorder for the transition is further emphasized when the $J_s(T)$ curves are rescaled by the $T=0$ value of the stiffness and by $T_c$, see Fig.\ \ref{fig0}. Here we finds a remarkable collapse of all the curves on each other, showing the complete irrelevance of disorder even away from criticality.
 According to the previous discussion, this follows from the fact that the leading temperature dependence below $T_c$ is due to the universal spin-wave suppression of the diamagnetic term, so that $T_c$ itself scales with $J_s(T=0)$.
 
 To simulate instead the effect of spatially-correlated disorder we perform Monte Carlo simulations with local maps of the couplings showing the "granular" structure mentioned in the introduction. These can be generated by a mean-field solution of the quantum $XY$ pseudo-spin 1/2 model in a transverse random field\cite{ma_prb85}:
\begin{equation}
\lb{hamps}
\mathcal{H}_{PS}\equiv -2\sum_i\xi_iS_i^z-2J\sum_{\langle i,j\rangle}\left(S^+_iS^-_j+h.c.\right).
\end{equation}
%
%
 \begin{figure*}[t]
\includegraphics[width=18cm]{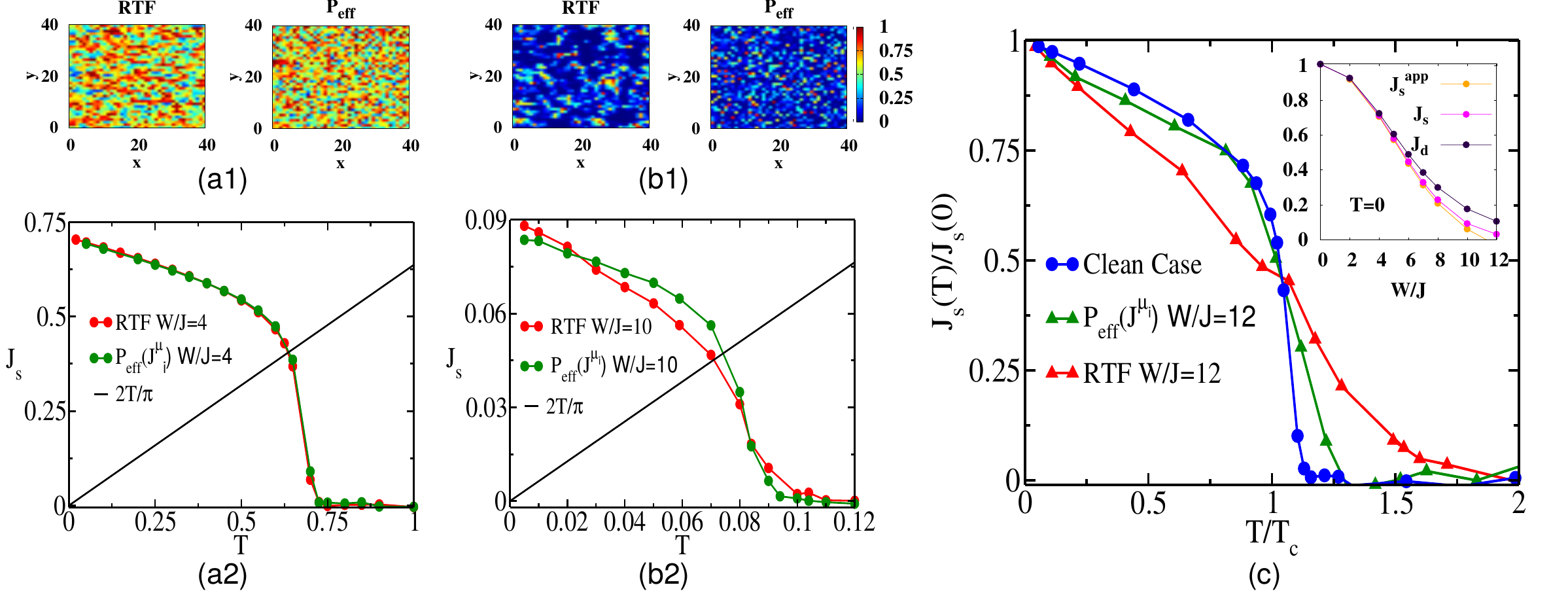}
\caption{Maps of the couplings $J_{i,i+x}$  (a1, b1) and superfluid stiffness (a2, b2) at disorder level $W/J=4$ and $W/J=10$. Here RTF and $P_{eff}$ denote the case of correlated and uncorrelated disorder, respectively. At weaker disorder (a2) the two curves overlap and the jump of $J_s$ is still sharp. At larger disorder (b2) the curve for the $RTF$ case starts to deviate from the usual trend, showing an almost symmetric smearing of the jump around the critical temperature. (c) Rescaled curves of the superfluid stiffness for the clean case, the uncorrelated $P_{eff}$ and correlated RTF disordered case at $W/J=12$. Despite the strong disorder the $P_{eff}$ curve shows only a small finite-size effect above $T_c$, while the RTF stiffness is dramatically modified above and below the transition. Inset: evolution with disorder of the zero-temperature value of  $J_s$, of $J_d$ for the RTF model and of the approximate result $J_s^{app}$ obtained from (\eqref{jsest}) }\label{fig1}
\end{figure*}
%
The model \pref{hamps} has been recently discussed in the literature within the context \cite{ioffe,lemarie_prb13,cea_prb14} of strongly-disordered superconductors. In the pseudo-spin language $S^{z} = \pm 1/2$ corresponds to a site occupied or unoccupied by a Cooper pair, while superconductivity corresponds to
a spontaneous in-plane magnetization, controlled by the coupling $J$. The random transverse field $\xi_i$, box distributed between $-W$ and $W$, mimics the effect of disorder, which tends to localize the Cooper pair on each site. At mean-field level\cite{suppl} the in-plane magnetization is $\langle S^x_i\rangle=\frac{1}{2}\sin\phi^{ps}_i$, where $\phi_i^{ps}$ is the angle that each pseudospin form with respect to the $z$ axis. While at small $W$  $\langle S^x_i\rangle\simeq 1/2$ everywhere, as $W/J$ increases the pseudo-spins partly orient out-of-the plane suppressing the in-plane component, i.e. the local value of the SC order parameter. 
In addition\cite{cea_prb14,suppl} the SC  in-plane phase fluctuations on top of this inhomogeneous SC ground state are controlled by a local stiffness  $J_{ij}=J\sin \phi^{ps}_i\sin\phi^{ps}_j$ which becomes itself a function of space. As discussed in Ref.s\ \cite{cea_prb14,lemarie_prb13} the local stiffness $J_{ij}$ is on average strongly suppressed by disorder, and it tends to form clusters of good SC regions embedded in a background with $J_{ij}\simeq 0$, as shown by the insets of Fig.\ \ref{fig1}a1, b1. 
Here we aim to explore how the BKT transition behaves in the presence of this non-trivial form of disorder. To simplify the approach we do not simulate the quantum model \pref{hamps}, but we derive from it a map of local couplings $J_{ij}$ to be injected in the model \pref{hamrtf}, where $\theta_i$ is the angle of the in-plane pseudospin component, i.e. the SC order parameter. The evolution of the stiffness computed by means of Monte Carlo simulations for increasing disorder level $W/J$ (with $J=1$) is shown in Fig.\ \ref{fig1}. To disentangle the effects of the spatial correlations of the couplings from the ones connected to their probability distribution,  we also compute for each disorder level the stiffness of the effective, uncorrelated distribution $P_{eff}$. This means that we assign the value $J_{ij}$ to each link by extracting it randomly from the same probability distribution $P_{eff}(J_{ij})$ which represents the RTF maps. In this case the SC state does not show any evident aggregation in real space, giving rise to standard, uncorrelated disorder, as it is already evident in the maps shown in Fig.\ \ref{fig1}a1,b1.The $T=0$ suppression of the stiffness is well captured by the approximated expression \pref{jsest} up to large $W/J$ values, see inset of Fig.\ \ref{fig1}c. With respect to the Gaussian case discussed above, here one has a large suppression of the diamagnetic term for increasing $W/J$, which explain the rapid suppression of the $T=0$ stiffness. On the other hand, up to $W/J=4$ the BKT transition preserves its character, and spatial correlations are irrelevant, as shown in Fig.\ \ref{fig1}a2. However at larger disorder the granularity of the SC state increases, and the superfluid-density jumps starts to be smeared out, see Fig.\ \ref{fig1}b2. Despite this, the same effect is not seen when spatial correlations disapper, as demonstrated by the case of $P_{eff}$. To compare the behavior at different disorder levels we shown in panel c the rescaled curves. At $W/J=12$ the probability distribution of the coupling is peaked at low values with very large tails\cite{ioffe,lemarie_prb13,suppl}. This has the only effect to increase slightly the finite-size effect, as one can see by comparing the curve for $P_{eff}$ with the clean case. However, the granular RTF model shows a definitively broader jump, which is symmetrically smeared out around $T_c$. 

%
 \begin{figure}[h!]
\includegraphics[width=5.5cm]{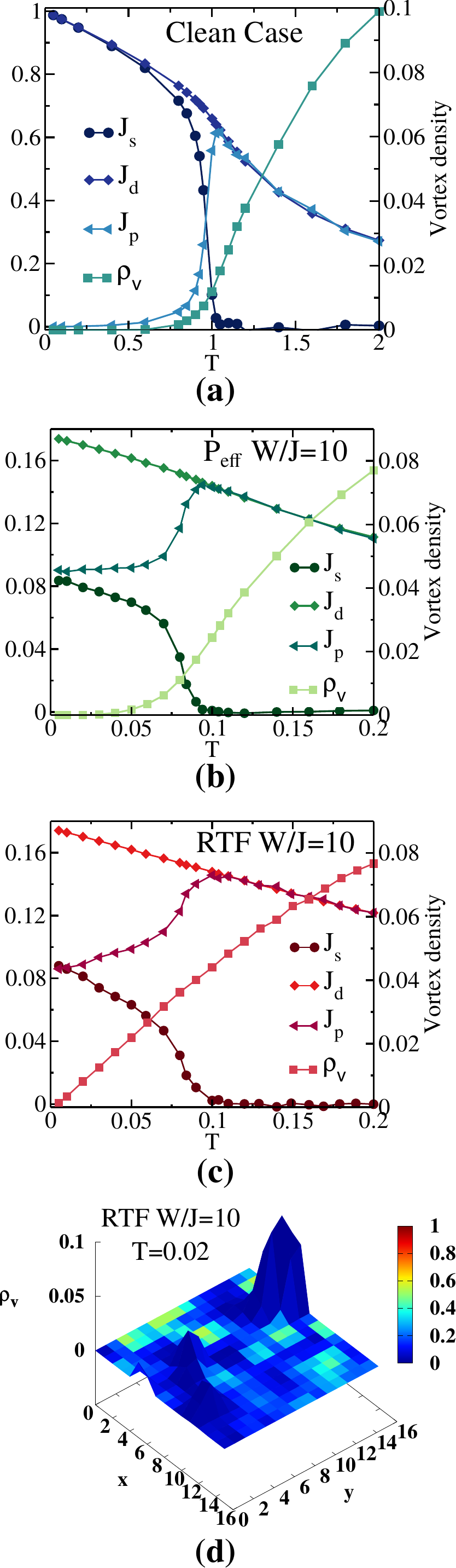}
\caption{Temperature dependence of the superfluid stiffness $J_s$, the diamagnetic term $J_d$, the paramagnetic term $J_p$ and the vortex pair density $\rho_v$ for three different cases: (a) Clean case, (b) Uncorrelated disorder $P_{eff}$ with $W/J=10$ and (c) Correlated disorder $RTF$ with $W/J=10$. (d) Local vortex density $\rho_V$ at $T=0.02$ superimposed to the colour map of the local stiffness $J_{ij}$ for the RTF model. }
\label{fig2}
\end{figure}

To get a deeper insight on the role of the spin-wave and vortex excitations contributing to the stiffness we show in Fig.\ \ref{fig2} the temperature evolution of the two separate diamagnetic \pref{jd} and paramagnetic \pref{jp} contributions, along with the average density $\rho_V$ of vortex pairs. This is defined by computing the local (positive or negative) vorticity of the phase around each square plaquette of the array. In Fig.\ \ref{fig2}a we show the results for the clean case. As discussed in Eq.\ \pref{jsest} above, spin-waves dominate the behavior of $J_d$ and $J_p$ at low temperatures. The vortex density is exponentially suppressed at low $T$ and it increases sharply at $T\simeq 0.9$, bringing up the paramagnetic contribution, which grows fast leading to $J_p=J_d$ at $T\simeq 1$. A similar trend in observed at $W/J=10$ for the $P_{eff}$ case, see Fig.\  \ref{fig2}b. Indeed, apart from the sizeable finite corrections \pref{jsest} to $J_d$ and $J_p$ at $T=0$, the thermal evolution of the various contributions is essentially the same: the vortex density has a fast increase only at $T\simeq 0.075$, where the universal jump is indeed expected (see Fig.\ \ref{fig1}b). The results change instead considerably for the RTF model, Fig.\ \ref{fig2}c.  In particular we observe an anomalous smooth increase of the paramagnetic response at low temperature, followed by a faster one around the temperature scale where the universal jump should be observed. This unconventional paramagnetic response explains the symmetric broadening of the transition observed in Fig.\ \ref{fig1}b. A second striking result is the almost linear increase of the vortex density in the whole temperature range. 

Even though a direct connection between the temperature evolution of $J_p$ and $\rho_V$ is not straightforward, nonetheless our results suggest that the anomalous temperature evolution of the stiffness at strong disorder originates from a change in the vortex nucleation mechanism, triggered by the fragmentation of the SC state. In the RTF model the formation of good SC islands is accompanied by the emergence of large clusters of bad SC regions, where the stiffness is very small. Vortices can then proliferate inside these regions, as shown in Fig.\ \ref{fig2}d where the local vorticity is superimposed to the colour map of the local stiffness. 
As a specular effect, when the vortex density increases, bringing the vortex pairs inside the good SC regions, the inhomogeneity of the SC state limit the entropic gain which usually triggers the vortex-pair dissociation in the homogeneous systems. In other words, vortices become "weakly" dangerous for the superfluid transport already below $T_c$, breaking down the usual "universal" balance between energetic and entropic gain, which is the hallmark of the BKT vortex-unbinding mechanism. Notice that the same anomalies below $T_c$ are observed by increasing the lattice size up to $L=256$\cite{suppl}, suggesting that the effects observed here are very different from the usual rounding of the stiffness above $T_c$ due to conventional finite-size effect.

In summary, we investigated by Monte Carlo simulation the evolution of the universal superfluid-density jump within a $XY$ model with random local couplings. We compared models with and without spatial correlations, focusing on the temperature dependence of the superfluid stiffness. When disorder lacks spatial structure it appears irrelevant not only for the jump at criticality, as expected, but also away from it. Indeed, by rescaling the stiffness to its $T=0$ value, suppressed by disorder, we observe a remarkable universal temperature dependence. This scenario changes considerably when disorder acquires spatial correlations, modelled here as a fragmentation of the SC state in good islands embedded in a bad SC background. In this case the superfluid-density jumps is considerably smeared out both above and below the temperature where the universal jump would be expected. This effect is attributed to a different mechanism for the vortex-antivortex pair generation due to the presence of large clusters of low-SC regions. Our results not only provide an explanation for the trends observed  experimentally in thin films of conventional
\cite{fiory_prb83,lemberger_prl00,armitage_prb07,armitage_prb11,kamlapure_apl10,mondal_bkt_prl11,goldman_prl12,yazdani_prl13} and unconventional\cite{lemberger_prb12} superconductors, but  they pave the way for the understanding of the topological excitations in gated 2D superconductors, where the inhomogeneity of the SC state is recently emerging\cite{biscaras_natmat13,bid_prb16,jespersen_prb16} as the hallmark of the field-induced electron doping.

\vspace{1cm}

\begin{acknowledgments}
We thank J. Lorenzana for useful discussions. L.B. acknowledges financial support by MIUR under projects FIRB-HybridNanoDev-RBFR1236VV, PRIN-RIDEIRON-2012X3YFZ2 and Premiali-2012 ABNANOTECH.
\end{acknowledgments}

\end{document}